\documentstyle[twocolumn,prl,aps]{revtex}
\input epsf

\begin{document}

\draft

\title{Role of the unstable directions in the equilibrium and 
aging dynamics of supercooled liquids.}

\author{Claudio Donati, Francesco Sciortino and Piero Tartaglia}

\address{Dipartimento di Fisica, Universita' di Roma
"La Sapienza" and Istituto Nazionale per la Fisica della Materia,
Piazzale Aldo Moro 2, I-00185, Roma, Italy}

\date{Revised version LM7663: \today}
\maketitle

\begin{abstract}

The connectivity of the potential energy landscape in supercooled
atomic liquids is investigated through the calculation of the
instantaneous normal modes spectrum and a detailed analysis of the
unstable directions in configuration space.  We confirm the hypothesis
that the mode-coupling critical temperature is the $T$ at which
the dynamics crosses over from free to activated exploration of
configuration space.  We also report the observed changes in the
local connectivity of configuration space sampled during aging,
following a temperature jump from a liquid to a glassy state.

\end{abstract}
\pacs{PACS numbers:  64.70.Pf, 61.20.Ja, 61.20.Lc}

Understanding the microscopic mechanism for the incredible slowing
down of the dynamics in supercooled glass forming liquids is
one of the hot topics in condensed matter physics.  In recent years, the
combined effort of high level experimental techniques\cite{exp},
computational analysis\cite{num,sri} and sophisticated theoretical
approaches\cite{goetze1,teo1,teo2} has provided an enormous amount of novel
information.  In particular, it appears more and more clearly that
in addition to the melting temperature 
and the calorimetric glass transition temperature, $T_g$,
another temperature plays a relevant
role.  This temperature, located between 1.2 and 2 $T_g$ depending on
the fragility of the liquid, signals a definitive change in the
microscopic processes leading to structural relaxation. Mode
Coupling Theory (MCT)\cite{goetze1} 
was first
in identifying the role of this cross-over temperature, $T_c$.
According to MCT, for $T \ge T_c$ the molecular dynamics is controlled
by the statistics of the orbits in phase space~\cite{orbits}, while
below $T_c$, the dynamics becomes controlled by phonon assisted
processes\cite{goetze1}.  Studies based on disordered mean field
$p$-spin models have also stressed the role of such a cross-over
temperature\cite{thirumalai}. 

Computer simulation studies of realistic models of liquids have 
addressed the issue of the structure of configuration space in
supercooled states~\cite{sri}. 
Two different techniques
have provided relevant information on phase space structure: the
instantaneous normal mode approach (INM)\cite{keyes1}, which focuses
on the properties of the finite temperature Hessian, allowing the
calculation of the curvature of the potential energy surface (PES)
along $3N$ independent directions and (ii) the inherent
structure (IS) approach\cite{stillinger}, which focuses on the
local minima of the potential energy. Different from mean
field models, computer simulation analysis provides a description
based on the system's potential energy, the free energy entering only
via the equilibrium set of analyzed configurations.

In principle, the INM approach should be well suited for detecting a
change in the structure of the PES visited above and below $T_c$.  A
plot of the $T$-dependence of the fraction of directions in
configuration space with negative (unstable) curvature, $f_u$, should
reveal the presence of $T_c$.  Unfortunately, as it was soon found
out\cite{keyes2}, anharmonic effects play a non negligible role
and several of the negative curvature directions are observed even in
crystalline states, where diffusivity is negligible.  A similar
situation is seen in p-spin models, where the number of negative
eigenvalues of the Hessian is not zero at $T_c$~\cite{biroli}.
Bembenek and Laird\cite{laird} suggested inspecting the energy profile
along the unstable directions to partition the negative eigenmodes in
shoulder modes ($sh$) (i.e. anharmonic effects) and double-well ($dw)$
modes (i.e. directions connecting different basins).  While for a
particular molecular system the fraction of double well modes $f_{dw}$
has been shown to go to zero close to $T_c$\cite{water}, for atomic
system, $f_{dw}$ is significantly different from zero even in
crystalline and glassy states\cite{laird}.  The existence of $dw$
directions in thermodynamic conditions where the diffusivity is zero,
i.e. in situations where the system is constrained in a well defined
basin, strongly suggests that the two minima joined by the unstable
double-well directions may lead to the same inherent structure
IS. Such possibility has been demonstrated very clearly in
Ref.\cite{gezelter}.

The aim of this Letter is to present a detailed evaluation of: (i) the
number of escape directions, $N_{escape}$, leading to a basin {\it
different} from the starting one; (ii) the number of distinct basins,
$N_{distinct}$, which are connected on average to each configuration
via a $dw$ direction.  This analysis, based on a
computationally demanding procedure, shows that indeed the PES regions
visited in thermal equilibrium above and below $T_c$ are clearly
different.  We also study the evolution of the sampled PES as a
function of time following a quench from above to below
$T_c$.
We show that, in analogy with the equilibrium
case, two qualitatively different dynamical regimes exist during
aging, related to different properties of the sampled PES.

The system we study is composed of a binary (80:20) mixture of 
$N=1000$ Lennard Jones atoms\cite{parameter}.
The dynamics of this system is well described by MCT, with a
critical temperature $T_c$ equal to $0.435$\cite{kob}.

We begin by considering the equilibrium case.  At each $T$, 
for each of the 50 analyzed 
configurations, we calculate the INM spectrum and --- by
rebuilding the potential energy profile along straight paths following
directions with negative curvature--- we classify the unstable modes
into $dw$ and $sh$. It is important to notice that 
the classification is done by studying 
the shape of the PES along one eigenvector,
i.e. beyond the point in configuration space where the eigenvector
was calculated. In principle, to identify a $dw$ or a $sh$ 
mode, one should follow a curvilinear path. Indeed 
the use of straight paths 
guarantees only the identification of the $dw$ modes whose 
one-dimensional saddle energy is close to the potential energy of the
system. As discussed in more length in \cite{discussion}
those $dw$ modes  are the ones relevant for describing motion in configuration 
space.

Figure \ref{fig:fdw} shows the $T$ dependence of the
unstable and $dw$ modes for the studied system. In agreement with the
analogous calculation of Ref.\cite{laird}, even at the lowest
temperature where equilibration is feasible within our computer
facilities, $f_{dw}$ is significantly different from zero.  To
estimate roughly the number of double wells which do not contribute
to diffusion, we have calculated the IS associated with the T=0.446
equilibrium configurations, and we have heated them back to various
temperatures below $T_c$. The corresponding $f_u$ and
$f_{dw}$ are also shown in Fig.\ref{fig:fdw}. Both these quantities
depend linearly on the temperature.  If we extrapolate $f_{dw}$ for
the equilibrium and for these non-diffusive cases, we find that the
two curves 
cross close to the MCT critical temperature $T_c$. Thus $T_c$
seems to be the temperature at which the number of
directions leading to different basins goes to zero, leaving only
local activated processes as residual channels for structural 
relaxation\cite{precisazione}.

To estimate in a less ambiguous way the number of different basins
which can be accessed from each configuration we apply the following
procedure: (i) calculate the $dw$ directions via INM calculation; (ii)
follow each straight $dw$ direction climbing over the potential 
energy barrier
and down on the other side until a new minimum is found; (iii) perform a
steepest descent path starting from the new
minimum found  along the $dw$ direction, as first suggested by Gezelter et
al.\cite{gezelter}; (iv) save the resulting $IS$ configuration; (v)
repeat (ii-iv) for each $dw$ mode.  This procedure produces a list of
$IS$s which can be reached from the initial configuration crossing a
one-dimensional $dw$.  We note that our procedure only guarantees that
the two starting configurations for the quenches are on different sides of
the double well, since --- as discussed in \cite{discussion} ---
there is some arbitrariness in the location of the two minima.

We calculate the relative distance $d_{ij}$

\begin{equation}
d_{ij}=\sqrt{\frac{1}{N}\sum_{l=1}^N (x_l^i-x^j_l)^2+(y_l^i-y^j_l)^2+
(z_l^i-z^j_l)^2}
\label{matrix}
\end{equation}
  
between all IS pairs in the list to determine the number of
{\it distinct} basins, $N_{distinct}$,  
connected to the starting configuration.  Here
$x_m^k$, $y_m^k$, and $z_m^k$ are the coordinates of the $m$-th particle
in the $k$-th IS of the list ($IS_k$).  
We also calculate the distance $d_{0i}$ ---
where $0$ indicates the IS associated to the starting configuration
--- to enumerate the number of escape directions, $N_{escape}$, leading
to a basin different from the original one ($IS_0$).

In Fig.~\ref{fig:doj} (upper panel) we show a plot of the distributions of the
distances $d_{0i}$ between the starting IS and the minima identified
with this procedure for different temperatures. At high temperatures,
the distribution has a single peak centered approximately at $d=0.3$.
For lower temperatures, the peak moves to the left and decreases in
height.  At the same time, a second but distinct peak, centered around
$d\approx 10^{-4}$, appears. 
Since our sample is composed of 
$1000$ particles, an average distance of the order of
$10^{-4}$ means that, also in the case that only one particle has a
different position between the two different configurations, this
particle has moved less that $0.003$ interparticle 
distance. Thus, we
consider the IS $i$ and $j$ as coincident if $d_{ij}<
10^{-2.5}$.

The upper panel of Fig.\ref{fig:NdvsT} reports the $T$-dependence of
the number of $dw$ directions $N_{dw}$, the number of escape
directions and the number of distinct basins which can be reached by
following a $dw$ direction.  We find that at high temperature nearly
all $dw$ directions leads to a different IS, thus fully contributing
to the diffusion process.  On lowering the temperature, a large
fraction of the directions leads to the same minimum, and, close to
$T_c$, almost all $dw$ directions lead to intra-basin motion.  A basin
change becomes a rare event.  Data in Fig.\ref{fig:NdvsT} support the
view that $T_c$ is associated to a change in the  PES
sampled by the system, and, consequently, in the type of
dynamics that the system experiences. Above $T_c$ the system can
change basins in configuration space by moving freely along an
accessible $dw$ direction, while below $T_c$ the thermally driven
exploration of configuration space favors the exploration of the
interior of the basins.  Diffusion in configuration space,
i.e. basin changes, requires an local activated process.

We next turn to the study of out-of-equilibrium dynamics.
Configurations are equilibrated at high temperature ($T=5.0$) and at
time $t_w=0$ are brought to low temperature $T_f=0.2$ by
instantaneously changing the control temperature of the thermostat.
Within 100 molecular dynamics steps, the kinetic energy of the system
reaches the value corresponding to $T_f$.  Configurations are recorded
for different waiting times $t_w$ after the quench. We then repeat the
same analysis done for the equilibrium configurations, as described
before.

Figure~\ref{fig:doj} (lower panel) shows the distribution of the
distances $d_{0i}$ between the minima $IS_0$ and $IS_i$. For short
$t_w$ the distribution has a single peak centered approximately at
$d_{0i}=0.3$, as found in the equilibrium configurations at high
temperature. As the system ages, the peak shifts to smaller distances
and a second distinct broad peak centered at approximately
$d_{0i}=5\times 10^{-4}$ appears.
 The observed behavior is similar to the
equilibrium one, on substituting $T$ with $t_w$.

The lower panel of Fig.~\ref{fig:NdvsT} shows $N_{dw}$, $N_{escape}$
and $N_{distinct}$ as functions of $t_w$.  At small $t_w$, following
a double well direction always brings to a basin that is distinct from
the original one, while at long times all $dw$ directions leads to
intra-basin motion. Again, the behavior is very reminiscent of the
equilibrium one, on substituting $T$ with $t_w$.  From these results,
we suggest that two qualitatively different dynamical processes
control the aging processes, respectively for short and long
$t_w$\cite{kobag}. For short waiting times the system is always close to saddle
points that allow it to move from one basin to a distinct one.  As the
system ages, it is confined to wells that are deeper and deeper, and
the number of basins that it can visit by following a $dw$ direction
decays to zero.  Thus, for long $t_w$ the dynamics of aging 
proceeds through local activated processes, that allow the system to pass
over potential energy barriers.  As shown in Fig.~\ref{fig:NdvsT},
beyond $t_w= 10^5$ the number of distinct basins is almost zero,
which, in analogy with the equilibrium results, supports the view that
a cross-over from MCT-like dynamics to hopping dynamics may take place
at a cross-over time during the aging process.

In summary, this Letter 
confirms the hypothesis that $T_c$ can be considered as the $T$
at which dynamics crosses over from the free-exploration to the
locally activated exploration of configuration space.  
In this respect, the
analogy between the slowing down of dynamics of liquids and the
slowing down of the dynamics in mean field $p$-spin model is
strengthened.  This Letter also shows that changes in dynamical
processes associated with changes in the explored local connectivity of
configuration space are observed even during the aging process, in
the time window accessed by molecular dynamics experiments.  
MCT-based models of aging\cite{latz} could be able to describe the early part
of the aging process --- i.e. the saddle-dominated dynamics ---
but may not be adequate for describing the locally activated
dynamical region.

We acknowledge financial support from the INFM PAIS 98, PRA 99 and 
{\it Iniziativa Calcolo Parallelo} and from MURST PRIN 98.  
We thank W. Kob, G. Parisi and T. Keyes for discussions.

\begin{figure}
\caption{$T$-dependence of $f_{u}$ (squares)  and $f_{dw}$ (circles).  
The
filled symbols are calculated from equilibrium configurations, the
empty symbols from out-of-equilibrium configurations below $T_c$ (see
text).}  
\label{fig:fdw}
\end{figure}

\begin{figure}
\caption{ a) Distribution of the distances $d \equiv d_{0i}$ between the IS $0$
and the $i$-th IS from equilibrium configurations at different
temperatures; b) same quantity for different
$t_w$ values
during the aging process after a temperature jump from $T_i=5.0$
to $T_f=0.2$.  }
\label{fig:doj}
\end{figure}

\begin{figure}
\caption{a) 
Number of double wells $N_{dw}$, of escape directions $N_{escape}$ and
of distinct basins $N_{distinct}$ for equilibrium configurations as a function of
$T$; b)Same quantities for different  $t_w$ values after a
temperature jump from $T_i=5.0$ to $T_f=0.2$. Time is measured in
molecular dynamics steps.}
\label{fig:NdvsT}
\end{figure}

\eject

\setcounter{figure}{0}

\begin{figure}
\hbox to\hsize{\epsfxsize=0.8\hsize\hfil
\epsfbox{fig1.epsi}\hfil}
\caption{C. Donati et al }
\end{figure}

\eject

\begin{figure}
\hbox to\hsize{\epsfxsize=0.8\hsize\hfil
\epsfbox{fig2a.epsi}\hfil}
\hbox to\hsize{\epsfxsize=0.8\hsize\hfil
\epsfbox{fig2b.epsi}\hfil}
\caption{C. Donati et al }
\end{figure}

\eject

\begin{figure}
\hbox to\hsize{\epsfxsize=0.8\hsize\hfil
\epsfbox{fig3a.epsi}\hfil}
\hbox to\hsize{\epsfxsize=0.8\hsize\hfil
\epsfbox{fig3b.epsi}\hfil}
\caption{C. Donati et al }
\end{figure}


\begin{references}
\bibitem{parameter} The interaction parameters
are given by:
$\epsilon_{AA}=1.0$, $\epsilon_{AB}=1.5$, $\epsilon_{BB}=0.5$,
$\sigma_{AA}=1.0$, $\sigma_{AB}=0.8$, $\sigma_{BB}=0.88$.
Lengths are defined in units of $\sigma_{AA}$, temperature $T$ in units of
$\epsilon_{AA}/k_B$, and time in units of $\sqrt{ m \sigma_{\rm AA}^2
/48\epsilon_{\rm AA}}$. Both atoms have unitary mass $m$. 
The simulations have been run for up to 
$5 \cdot 10^6$ MD steps in the NVT ensemble (Nose-Hoover thermostat).
One MD step is 0.02.
\bibitem{precisazione}
Such estimate of the 
free-exploration to activated dynamics  cross-over temperature
could be biased by the arbitrary choice of the $IS$
configuration (the $T=0.446$ configurations).
\end{references}
\end{document}